\documentclass{pos}
\usepackage{amsmath}

\title{NNLO corrections to charmless hadronic B decays}

\ShortTitle{NNLO corrections to charmless hadronic B decays}

\author{\speaker{Guido Bell}%
        \thanks{I would like to thank Volker Pilipp for a fruitful
          collaboration on part of the subjects presented in this
          article and the organizers of EFT09 for creating a pleasant
          and stimulating workshop atmosphere. This work was supported
          by the DFG Sonderforschungsbereich/Trans\-regio 9.}   
        \\
        Institut f\"ur Theoretische Teilchenphysik,
        Universit\"at Karlsruhe,
        76128 Karlsruhe, Germany\\
        E-mail: \email{bell@particle.uni-karlsruhe.de}}

\abstract{
We report on the status of the perturbative calculation in the QCD
factorization framework to charmless hadronic $B$ meson decays and
present a compilation of available NNLO results.} 

\FullConference{
  International Workshop on Effective Field Theories: from the pion to
  the upsilon\\
  February 2-6 2009\\
  Valencia, Spain}

\def\calO{\mathcal{O}}
\def\as{\alpha_s}
\def\LQCD{\Lambda_\text{QCD}}
\def\SCETI{\ensuremath{\text{SCET}_\text{I}}}
\def\SCETII{\ensuremath{\text{SCET}_\text{II}}}
\def\Br{\text{Br}}

\begin{document}

\section{Introduction}

$B$ meson decays into a pair of light mesons are mediated by rare
flavour-changing $b\to u$ (''tree'') or loop-induced $b\to d,s$
(''penguin'') quark transitions. The plethora of hadronic two-body final
states, consisting of e.g.~$\pi,\rho, K, K^*,\eta$ or $\phi$ mesons,
opens a particularly rich laboratory for testing the CKM paradigm of
flavour mixing and CP violation. 

In order to exploit the wealth of data that has been collected at the
$B$ physics experiments, a quantitative control of the
strong-interaction effects is essential. QCD factorization
(QCDF)~\cite{QCDF} is a systematic framework to compute the hadronic
matrix elements from first principles. It is based on the statement that
the matrix elements of the operators in the weak effective Hamiltonian
factorize in the heavy quark limit $m_b\gg\LQCD$ according to 
\begin{eqnarray}
\label{eq:fact}
\langle M_1 M_2 | Q_i | \bar{B} \rangle
&\;\simeq\; &
F^{B M_1}(0) \; f_{M_2}
\int du \;T_{i}^I(u) \; \phi_{M_2}(u)
\nonumber \\
&&
+ \; \hat{f}_{B} \; f_{M_1} \; f_{M_2}
\int d\omega dv du \; T_{i}^{II}(\omega,v,u)
\; \phi_B(\omega) \; \phi_{M_1}(v) \;  \phi_{M_2}(u),
\end{eqnarray}
where the non-perturbative effects are confined to some
process-independent hadronic parameters such as decay constants $f_M$,
light-cone distribution amplitudes $\phi_M$ and a heavy-to-light form
factor $F^{B M}(0)$ at maximum recoil. The short-distance
hard-scattering kernels $T_i^{I,II}$, on the other hand, are
perturbatively calculable and currently being worked out to
next-to-next-to-leading order (NNLO), i.e.~at
$\calO(\as^2)$~\cite{NNLO:T2:tree,NNLO:T2:peng,NNLO:T2:tree:Volker,NNLO:T1:tree}.
Here we report on the status of the perturbative calculation, which we
divide into two parts: vertex corrections ($T_i^{I}$) and spectator
scattering ($T_i^{II}$).

\section{Spectator scattering}
\label{sec:spec}

\begin{figure}[b]
\centerline{
\includegraphics[height=2.25cm]{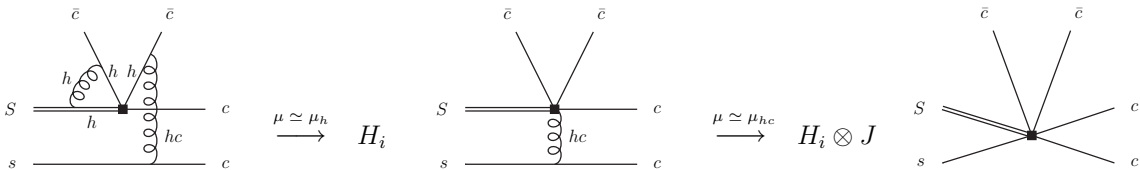}}
\caption{Two-step matching procedure
  QCD~$\to$~$\SCETI$(hc,c,s)~$\to$~$\SCETII$(c,s) of the spectator
  scattering contribution. $S$ ($s$) denotes a soft heavy (light) quark,
  $c$ ($\bar c$) a collinear quark in the direction of the meson $M_1$
  ($M_2$) and $h$ ($hc$) refers to virtual hard (hard-collinear) modes.}
\label{fig:T2}
\end{figure}

We start with the class of short-distance interactions that involves the
spectator quark of the decaying $B$ meson. Technically, the calculation
of the kernels $T_i^{II}$ amounts at the considered order $\alpha_s^2$
to a 1-loop calculation with six external legs, cf.~the left diagram of
Figure~\ref{fig:T2}. The calculation is complicated by the fact that the
interactions between the soft spectator and the energetic (collinear)
particles in the final state induce a new (virtual) degree of
freedom. These so-called hard-collinear or jet modes describe
configurations of energetic massless particles with virtualities
$\mu_{hc}^2\simeq m_b\LQCD$, which is in between the hard scale
$\mu_h\simeq m_b$ and the hadronic scale $\LQCD$. 

The decomposition in (\ref{eq:fact}) relies on a perturbative treatment
of these hard-collinear modes (it is usually assumed that
$\mu_{hc}\simeq1.5$~GeV). As the kernels $T_i^{II}$ contain the effects
from two short-distance modes with different virtualities, the
calculation becomes most transparent when it is organized as a two-step
matching calculation between QCD and soft-collinear effective theory
(SCET)~\cite{SCET}, where the perturbative degrees of freedom are
subsequently integrated out. 

The matching calculation is illustrated in Figure~\ref{fig:T2}: QCD is
first matched at $\mu\simeq \mu_h$ onto an effective theory called
$\SCETI$(hc,c,s) and then at a lower scale $\mu\simeq \mu_{hc}$ onto
$\SCETII$(c,s). The respective matching coefficients $H_i$ ($J$) encode
the effects of the hard (hard-collinear) modes, while the hadronic
matrix element of the remnant non-local operator in $\SCETII$(c,s)
yields the factorized light-cone distribution amplitudes $\phi_M$ of the
three mesons. The hard-scattering kernels $T_i^{II}$ finally follow from
a convolution of the perturbative coefficient functions
\begin{eqnarray}
\label{eq:fact:T2}
T_{i}^{II}(\omega,v,u) \;=\;
\int dz \;\,  J(\omega,v,z) \; H_i(z,u).
\end{eqnarray}
From Figure~\ref{fig:T2} it is evident that the spectator scattering
mechanism requires at least one perturbative gluon exchange with
$J=\calO(\as)$ and $H_i=\calO(1)$ (at tree level). It has been pointed
out in~\cite{Bauer:2004tj} that the same jet function $J$ also enters
the factorization formula for heavy-to-light form
factors~\cite{fact:formfactor}. As the $\calO(\as^2)$ corrections to $J$
have already been worked out in this context~\cite{NNLO:T2:jet}, the
NNLO calculation of the kernels $T_i^{II}$ reduces to the computation of
the $\calO(\as)$ terms of the hard coefficient functions $H_i$.  This
programme has recently been completed: the corrections for the
topological tree amplitudes have been computed in~\cite{NNLO:T2:tree}
and the ones for the QCD and electroweak penguin amplitudes
in~\cite{NNLO:T2:peng}. 

The work in~\cite{NNLO:T2:tree:Volker} follows an alternative approach
to compute the kernels $T_i^{II}$ (for the tree amplitudes). In this
work the calculation has been performed in pure QCD by expanding the
1-loop diagrams to leading power in $1/m_b$. The calculation thus yields
directly the convolution from (\ref{eq:fact:T2}) without disentangling
hard and hard-collinear effects. The findings
of~\cite{NNLO:T2:tree:Volker} agree with those
from~\cite{NNLO:T2:tree,NNLO:T2:jet}, which demonstrates the equivalence
of the diagrammatical approach (QCDF) and the effective field theory
formulation (SCET) at a non-trivial fixed order in perturbation theory.

\section{Vertex corrections}

\begin{figure}[b]
\centerline{
\includegraphics[height=1.65cm]{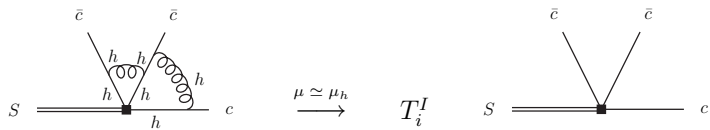}}
\caption{Matching of the 2-loop vertex corrections (the spectator quark
  is irrelevant for this contribution and not drawn). The different
  configurations are described in the caption of Figure~1.} 
\label{fig:T1}
\end{figure}

The calculation of the kernels $T_i^{I}$ is conceptually simpler, since
the complicated interplay of soft and collinear dynamics, which induces
the hard-collinear degrees of freedom, is absent in this case. The
calculation is, however, technically demanding as it amounts to a 2-loop
calculation with four external legs, cf.~Figure~\ref{fig:T1}.

The 2-loop corrections to the kernels $T_i^{I}$ for the topological tree
amplitudes have been computed in~\cite{NNLO:T1:tree}. The work makes use
of a couple of advanced techniques, which are widely applied in
multi-loop calculations. First and foremost the calculation is based on
an automatized reduction algorithm, which uses integration-by-parts
techniques~\cite{IBP} to relate the entire 2-loop calculation to an
irreducible set of scalar Master Integrals (MIs). The actual number of
36 MIs turns out to be large and the presence of soft and collinear
infrared divergences in intermediate steps of the calculation implies
that the MIs are required to up to five coefficients in the
$\varepsilon$-expansion, which makes the calculation somewhat involved
($\varepsilon=(4-d)/2$ in dimensional regularization).

The whole set of MIs has been computed in~\cite{NNLO:T1:tree}. Several
sophisticated techniques, among these the method of differential
equations~\cite{DiffEqs}, the formalism of Harmonic
Poly\-logarithms~\cite{HPLs} and Mellin-Barnes
techniques~\cite{MellinBarnes},  have been applied to derive analytical
expressions for all MIs. The results have further been checked
numerically with the method of sector
decomposition~\cite{SecDecomp}. About half of these MIs have later been
confirmed and applied in a calculation for inclusive semileptonic $B\to
X_u\ell\nu$ decays by various groups~\cite{BtoXuellnu}. One particular
coefficient of one of the most complicated MIs has first been given
numerically in~\cite{NNLO:T1:tree,BtoXuellnu} and later been rederived
in an analytical form~\cite{Huber:2009se}.

Apart from the calculation of 2-loop integrals, the NNLO calculation of
the kernels $T_i^{I}$ reveals further technical difficulties. The
factorization formula (\ref{eq:fact}) implies in particular that the
(infrared divergent) contributions, which belong to the formal expansion
of the non-perturbative objects ($F^{B M}, f_{M}, \phi_{M}$), have to be
subtracted from the partonic calculation. In dimensional regularization
this matching calculation involves non-physical (evanescent)
operators. In order to assure that these operators disappear from the
final factorization formula, their renormalized matrix elements have to
be made to vanish. The specific pattern of required infrared
subtractions is particularly complicated  in the calculation of the
colour-suppressed tree amplitude, which involves a Fierz-evanescent
operator already at tree level. 

The calculation itself provides a couple of stringent
cross-checks. Whereas individual 2-loop diagrams contain up to
$1/\varepsilon^4$ (soft and collinear) infrared divergences, the kernels
$T_i^I$ turn out (as predicted) to be free of any singularities, which
follows after an intricate subtraction procedure of ultraviolet and
infra\-red divergences. Moreover, somewhat subtle renormalization scheme
dependent contributions enter at the 2-loop level scale-dependent terms,
which can also be checked. Another important check of the scheme
independence of the NNLO results would be possible given a
Fierz-symmetric renormalization scheme of evanescent
operators. Unfortunately, such a scheme has not yet been worked out at
NNLO and therefore this final check is currently lacking.

In view of the complexity of the considered calculation, it is desirable
that there are (at least) two independent calculations of the kernels
$T_i^I$ (as for the spectator scattering kernels $T_i^{II}$). One
on-going calculation of the tree amplitudes is close to
finish~\cite{BHL}. The only missing ingredient of the NNLO calculation
then consists in the calculation of the kernels $T_i^I$ for the penguin
amplitudes~\cite{NNLO:T1:peng}.

\section{Compilation of NNLO results}

\subsection{Conceptual issues}

First of all it is worth noting that factorization has been found in all
available NNLO contributions to work technically, i.e.~soft and
collinear infrared singularities factorize as predicted at NNLO and the
resulting convolution integrals are finite (in particular free of
endpoint singularities). 

Another important conceptual point deals with the question if the
dynamical hard-collinear scale $\mu_{hc}\simeq \sqrt{m_b\LQCD}$ should
be treated as a perturbative scale, an issue that has been controversial
in the literature~\cite{Bauer:2004tj,SCETvsQCDF}. The question is
related to the perturbative expansion of the spectator scattering
mechanism discussed in Section~\ref{sec:spec}. As this contribution
starts at $\calO(\as)$, the NNLO terms constitute the first radiative
corrections to this mechanism. The factorization of this contribution
within SCET, as outlined in Section~\ref{sec:spec}, further allows to
systematically disentangle the effects from the scales $\mu_h$ and
$\mu_{hc}$ and in addition to resum logarithms $\ln \mu_h/\mu_{hc}$ via
renormalization group techniques. The explicit NNLO
results~\cite{NNLO:T2:tree,NNLO:T2:peng,NNLO:T2:tree:Volker,NNLO:T2:jet}
do not show any sign of abnormally large corrections and the remnant
dependence on the scales $\mu_h$ and $\mu_{hc}$ turns out to be
well-behaved. The NNLO calculation thus suggests that the expansion in
$\as(\mu_{hc})$ is under control.

\subsection{Tree amplitudes}

One important subset of hard-scattering kernels, which specify the
topological tree amplitudes, are by now completely determined to
NNLO~\cite{NNLO:T2:tree,NNLO:T2:tree:Volker,NNLO:T1:tree}. It is
interesting to illustrate the structure of the perturbative expansion at
the amplitude level. Using the input parameters specified
in~\cite{GBVP}, the colour-allowed ($\alpha_1$) and colour-suppressed
($\alpha_2$) tree amplitudes in the $B\to\pi\pi$ channels become 
\begin{align}
\alpha_1(\pi\pi)& \,=\,~~
                        1.008 \big{|}_{V^{(0)}}
         + \big[ 0.025 + 0.010i \big]_{V^{(1)}}
         + \big[ 0.027 + 0.032i \big]_{V^{(2)}}
\nonumber\\
&\hspace{4.5mm}
                      - 0.012 \big{|}_{S^{(1)}}
         - \big[ 0.021 + 0.015i \big]_{S^{(2)}}
                      - 0.014 \big{|}_{P}
\nonumber\\
& \,=\,~~
         1.013^{+0.023}_{-0.036} + (+0.027^{+0.025}_{-0.022})i,
\nonumber\\
\alpha_2(\pi\pi)& \,=\,~~
                        0.223 \big{|}_{V^{(0)}}
         - \big[ 0.174 + 0.075i \big]_{V^{(1)}}
         - \big[ 0.032 + 0.051i \big]_{V^{(2)}}
\nonumber\\
&\hspace{4.5mm}
                      + 0.090 \big{|}_{S^{(1)}}
         + \big[ 0.034 + 0.025i \big]_{S^{(2)}}
                      + 0.055 \big{|}_{P}
\nonumber\\
& \,=\,~~
         0.195^{+0.134}_{-0.089} + (-0.101^{+0.061}_{-0.063})i,
\label{eq:trees}
\end{align}
where the various terms of the perturbative expansion have been denoted
by $V^{(0)}$ corresponding to the tree level contribution, $V^{(1)}$
(1-loop vertex corrections) and $S^{(1)}$ (tree level spectator
scattering) to the NLO corrections and $V^{(2)}$ (2-loop vertex
corrections) and $S^{(2)}$ (1-loop spectator scattering) to the new NNLO
terms, whereas the numbers denoted by $P$ give an estimate of power
corrections $\sim1/m_b$ that are related to subleading twist wave
functions of the pions. 

From (\ref{eq:trees}) it is obvious that the NNLO corrections are
particularly important for the imaginary parts of the amplitudes and
hence for strong phases and direct CP asymmetries, which are first
generated at $\calO(\as)$. The new corrections are in some cases found
to exceed the NLO terms, which can be explained by a numerical
enhancement from the Wilson coefficients. In absolute terms, however,
the new corrections are small and perturbation theory seems to be
well-behaved. 

The uncertainties of the NNLO prediction for the colour-allowed tree
amplitude $\alpha_1$ are at a satisfactory level of few percent. The
colour-suppressed amplitude $\alpha_2$, on the other hand, suffers from
substantial uncertainties, which can be traced back to the strong
cancellation between the terms denoted by $V^{(0)}$, $V^{(1)}$ and
$V^{(2)}$. This makes the real part of $\alpha_2$ particularly sensitive
to the spectator scattering mechanism, which is proportional to the
hadronic ratio $f_{\pi}\hat{f}_B/\lambda_BF_+^{B \pi}(0)$. The poor
knowledge of the $B$ meson parameter $1/\lambda_B\equiv\int_0^\infty
d\omega/\omega\; \phi_B(\omega)$ in particular makes the theoretical
prediction of $\alpha_2$ rather uncertain, which calls for further
theoretical progress from non-perturbative methods  (the numbers are
given for $\lambda_B=(400\pm150)$MeV).

\subsection{The approximate tree decays $B^-\to\pi^-\pi^0/\rho^-\rho^0$}

Isospin symmetry implies that the decays $B^-\to\pi^-\pi^0/\rho^-\rho^0$
are free of QCD penguin contributions (they depend, however, on small
electroweak penguin amplitudes). As they do not receive contributions
from weak annihilation either, which constitutes an important class of
non-factorizable power corrections to the factorization formula
(\ref{eq:fact}), their branching ratios are particularly suited to test
the strong inter\-action dynamics of the tree amplitudes. Treating the
electroweak penguin amplitudes in the NLO
approximation~\cite{QCDF,QCDF:VV}, the QCDF prediction
becomes~\cite{GBVP} 
\begin{align}
%
10^6\,\Br(B^-\to\pi^-\pi^0) &  \;=\;
6.22\,^{+1.14}_{-1.05}\,^{+2.03}_{-1.65}\,^{+0.16}_{-0.18}\,^{+0.43}_{-0.42}
&&
(5.59^{+0.41}_{-0.40}),
\nonumber\\
10^6\,\Br(B^-\to\rho_L^-\rho_L^0) &  \;=\;
21.0\,^{+3.9}_{-3.5}\,^{+7.4}_{-6.1}\,^{+0.5}_{-0.7}\,^{+1.5}_{-1.4}
&&
(22.5^{+1.9}_{-1.9}),
\end{align}
which is in good agreement with the experimental values given in
parentheses. Here $L$ refers to the longitudinal polarization and the
uncertainties are, in order, due to CKM parameters, hadronic parameters,
higher order perturbative corrections and power corrections. The
theoretical prediction depends, however, strongly on the input values
for $|V_{ub}|$ and the form factors $F_+^{B\pi}(0)$ and
$A_0^{B\rho}(0)$. 

The theoretical prediction can be made independent of these input
parameters by normal\-izing the hadronic decay rates to the differential
semileptonic decay rates at maximum recoil. The ratio 
\begin{align}
\frac{\Gamma(B^- \to \pi^- \pi^0)}
{d\Gamma(\bar{B}^0\to \pi^+\ell^-\bar{\nu}_l)/dq^2|_{q^2=0}}
\;\simeq\;
3\pi^2 f_\pi^2 |V_{ud}|^2
|\alpha_1(\pi\pi)+\alpha_2(\pi\pi)|^2
\label{eq:Rpi}
\end{align}
provides an exceptionally clean probe of the QCD dynamics of the tree
amplitudes~\cite{facttest}. The NNLO prediction for this ratio
$(0.70^{+0.12}_{-0.08})\text{GeV}^2$ compares again well to the
experimental value $(0.81^{+0.14}_{-0.14})\text{GeV}^2$, which strongly
supports the factorization assumption~\cite{GBVP}. It would, however, be
interesting to see if the tendency between the experimental and the
theoretical value is reproduced in the $\rho$-sector. As the
semileptonic $B\to\rho\ell\nu$ decay spectrum has not yet been measured
precisely, one may instead consider a ratio of two hadronic decay rates 
\begin{align}
\frac{\Gamma(B^- \to \rho_L^- \rho_L^0)}
{\Gamma(\bar{B}^0 \to \rho_L^+ \rho_L^-)}
\;\simeq\;
\frac{|\alpha_1(\rho_L\rho_L)+\alpha_2(\rho_L\rho_L)|^2}{2|\alpha_1(\rho_L\rho_L)|^2},
\end{align}
which, in contrast to (\ref{eq:Rpi}), receives corrections from QCD
penguin amplitudes and weak annihilation. The NNLO prediction for this
ratio $(0.65^{+0.16}_{-0.11})$ is again found to be somewhat smaller
than the experimental value $(0.89^{+0.14}_{-0.14})$, which may be
considered as a hint at somewhat enhanced colour-suppressed amplitudes
that could be realized in QCDF by a smaller value of the $B$ meson
parameter $\lambda_B\simeq~250$MeV~\cite{GBVP}.

\section{Conclusions}

The NNLO calculation for charmless hadronic $B$ meson decays is
particularly important in respect of direct CP asymmetries that are
first generated at $\calO(\as)$. It opens in particular a new mechanism
from spectator scattering that induces strong phases and settles some
conceptual aspects that bring the factorization framework onto a more
rigorous footing. 

Whereas the topological tree amplitudes are by now completely determined
to NNLO, the computation of the penguin amplitudes is to date still
incomplete. Further improvements on the calculation require a better
understanding of power corrections, in particular on the role of the
phenomenologically important scalar penguin amplitude, and more precise
determinations of hadronic input parameters.

\end{document}